\documentclass[preprint,12pt]{elsarticle}

\usepackage{amsfonts,amsmath,amssymb,bm,a4wide,graphicx,makeidx,multicol,color}
\journal{Physics Letters B}
\begin{document}
\begin{frontmatter}
\title{The effect of plasmon mass on spin light of neutrino \\
in dense matter}

\author{A.Grigoriev$^{a}$,\ A.Lokhov$^{b}$,\
        A.Studenikin$^{b,c}$ \footnote{studenik@srd.sinp.msu.ru }, \
        A.Ternov$^{d}$}
\address{
    $^{a}$Skobeltsyn Institute of Nuclear Physics, \ Moscow State
        University, 119991 Moscow, Russia

    $^{b}$Department of Theoretical Physics, \ Moscow State University,
        119991 Moscow, Russia

    $^{c}$Joint Institute for Nuclear Research, 141980 Dubna, Russia

    $^{d}$Department of Theoretical Physics, \ Moscow Institute for Physics and Technology, 141700
Dolgoprudny, Russia}

\begin{abstract}
We develop the theory of spin light of neutrino in matter ($SL\nu$)
and include the effect of plasma influence on the emitted photon. We
use the special technique based on exact solutions of particles wave
equations in matter to perform all the relevant calculations, and
track how the plasmon mass enters the process characteristics
including the neutrino energy spectrum, $SL\nu$ rate and power. The
new feature it induces is the existence of the process threshold for
which we have found the exact expression and the dependence of the
rate and power on this threshold condition. The $SL\nu$ spatial
distribution accounting for the above effects has been also obtained.
These results might be of interest in connection with the recently
reported hints of ultra-high energy neutrinos $E = 1 \div 10 $ PeV
observed by IceCube.
\end{abstract}

\end{frontmatter}

\section{Introduction}

Neutrino physics in matter and external electromagnetic fields is a
rather longstanding research field nevertheless still having advances
and providing some interesting predictions for various phenomena. A
broad spectrum of issues here are connected with possible
electromagnetic properties of neutrino (for more details refer to
\cite{GiuStuYF09}). The recent studies of neutrino electromagnetic
properties revealed a new mechanism of electromagnetic radiation by a
neutrino propagating in dense matter that has been proposed in
\cite{LobStuPLB03}. This type of electromagnetic radiation was called
the spin light of neutrino in matter ($SL\nu$). In a quasi-classical
treatment this radiation originates due to neutrino electromagnetic
moment precession in dense background matter. The quantum theory of
this phenomena has been developed in \cite{StuTerPLB05,Lob05}.

A new convenient and elegant way of description of neutrino
interaction processes in matter has been proposed and developed in a
series of papers \cite{StuTerPLB05,StuJPA_06_08} (see also
\cite{Lob05}). The elaborated method is based on the use of the exact
solutions of the modified Dirac equation for a neutrino (the method
of exact solutions) in matter. The method was used to describe the
phenomenon of the spin light of neutrino in matter within quantum
approach. Note that the most recent development of the method in
application to studies of neutrino motion in different extended
environments can be found in \cite{BalPopStu11}.

The discussed method is based on the use of the modified Dirac
equations for the particles wave functions, in which the
correspondent effective potentials accounting for matter influence on
the particles are included. It is similar to the Furry representation
\cite{FurPR51} in quantum electrodynamics, widely used for
description of particles interactions in the presence of external
electromagnetic fields. In this technique, the evolution operator
$U_{F}(t_1, t_2)$, which determines the matrix element of the
process, is represented in the usual form
\begin{equation}
U_{F} (t_1, t_2)= T exp \bigg[-i \int
\limits_{t_1}^{t_2}j_{\mu}(x) A^{\mu}{d}x  \bigg],
\end{equation}
where $A_{\mu}(x)$ is the quantized part of the potential
corresponding to the radiation field, which is accounted within
the perturbation-series techniques. At the same time, the electron
(a charged particle) current is represented in the form
\begin{equation}
j_{\mu}(x)={e \over 2}\big[\overline \Psi_e \gamma _{\mu}, \Psi_e
\big],
\end{equation}
where $\Psi_e$ are the exact solutions of the Dirac equation for
the electron in the presence of external electromagnetic field
given by the classical non-quantized potential $A_{\mu}^{ext}(x)$:
\begin{equation}\label{D_eq_QED}
\left\{ \gamma^{\mu}\big(i\partial_{\mu} -eA_{\mu}^{cl}(x)\big) -
m_e \right\}\Psi_e (x)=0.
\end{equation}

By analogy with the above-mentioned Furry representation one can
introduce effective potentials for the impact of the background
matter on the propagation of the particles.  In the presence of
matter a neutrino dispersion relation is modified \cite{ManNie}, in
particular, it has a minimum at nonzero momentum \cite{ChangiDr}. As
it was shown in \cite{ManNie}, the standard result for the MSW effect
can be derived using a modified Dirac equation for the neutrino wave
function with a matter potential proportional to the density being
added. The problem of a neutrino mass generation in different media
was studied \cite{OraevskyiDr} with use of modified Dirac equations
for neutrinos. On the same basis spontaneous neutrino-pair creation
in matter was also studied \cite{LoebiDr}.

In this letter we make a reasonable step towards the completeness of
the physical picture and consider the question of the plasmon mass
influence on $SL\nu$. The $SL\nu$ is a process of photon emission in
neutrino transition between different helicity states in matter. As
it has been shown \cite{StuTerPLB05}, in the relativistic regime
$SL\nu$ mechanism could provide up to one half of the initial
neutrino energy transition to the emitted radiation. It was also
shown that the $SL\nu$ provides the spin polarization effect of
neutrino beam moving in matter (similarly to the well-known effect of
the electron spin self-polarization in synchrotron radiation
\cite{SokTer63}). The estimations already performed in
\cite{KuznMikh07} indicated that the plasmon has a considerable mass
that can affect the physics of the process. To see how the plasmon
mass enters the $SL\nu$ quantities we appeal to the method of exact
solutions and carry out all the computations relevant to $SL\nu$ in
an exact way and track how the plasmon mass enters the process
characteristics including the neutrino energy spectrum, $SL\nu$ rate
and power with particular focus on the spatial distribution.

\section{Kinematics of the process}

First we briefly consider the external conditions in which the
$SL\nu$ process is possible. Suppose the matter density is close to
that of nuclear matter. In astrophysics it could be found, for
example, in neutron stars with density span up to $10^{38-41}
cm^{-3}$ \cite{BelvedNPA12}). As it has been shown in
(\cite{StuTerPLB05}) in the case when neutrons are the main component
in the background matter, in fact, an antineutrino (not neutrino)
produce the $SL\nu$. Nevertheless we shall call it neutrino for
convenience (the relevant formulae remain the same).

In turn, the plasmon effects emerge only due to the electron matter
component which is considerably less dense in typical matter of a
neutron star. For definiteness, in what follows, in respect of
densities we assume the condition
\begin{equation}\label{n_e&n_n}
    n_e\simeq 0.1 n_n \,.
\end{equation}
Due to the smallness of the charged fraction we neglect its influence
on neutrinos and account below only for the neutron matter component
in any quantities associated with neutrino.

For the description of neutrino propagation in matter we use the
exact solutions of the modified Dirac equation for the neutrino field
in matter \cite{StuTerPLB05}
\begin{equation}
    \left\{i\gamma_{\mu}\partial^{\mu}-\frac{1}{2}\gamma_{\mu}(1+\gamma^{5})f^{\mu}-m_{\nu}\right\}\Psi(x)=0,
\label{eq:dirac}
\end{equation}
where in the case of the particle motion through the non-moving and
unpolarized neutron matter $f^{\mu}=-G_{F}/\sqrt{2} \
(n_n,\textbf{0})$. With that equation (\ref{eq:dirac}) has plane-wave
solution determined by the 4-momentum $p^{\mu}=(E,\bf{p})$ and
quantum numbers of the helicity $s=\pm 1$ and the sign of energy
$\varepsilon=\pm 1$. The corresponding neutrino energy is given by
\begin{equation}
    E=\sqrt{(p+s\tilde{n})^{2}+m_{\nu}^{2}}+\tilde{n}, \ \ \tilde{n} =
\frac{1}{2\sqrt{2}}G_F n_n.
 \label{eq:dispersion}
\end{equation}
The neutrino-photon coupling is described by the usual dipole
electromagnetic vertex ${\bf \Gamma}=i\omega\big\{\big[{\bf \Sigma}
\times {\bm \varkappa}\big]+i\gamma^{5}{\bf \Sigma}\big\}$, for
details see \cite{StuTerPLB05,StuJPA_06_08}. From the energy-momentum
conservation one obtains
\begin{equation}
    E=E^{\prime}+\omega; \ \ \bf{p}=\bf{p^{\prime}}+\bf{k},
\label{eq:conservation}
\end{equation}
where the primed values correspond to the outgoing neutrino. One has
to account for the plasma dispersion law in the form of
\begin{equation}\label{photon dispersion}
    \omega=\sqrt{k^2+m^2_{\gamma}}, \quad m_{\gamma}=\sqrt{2\alpha}(3\sqrt{\pi}n_e)^{1/3}.
\end{equation}

We consider the relativistic energy limit and $s=1, s'=-1$ that is the most relevant case for applications. From the energy-momentum conservation (4) we obtained that
\begin{equation}\label{kinematics}
    \omega\sqrt{(p+\widetilde{n})^{2}+m_{\nu}^{2}}=\widetilde{n}(p^{\prime}+p)+pk\cos\theta+\frac{m^2_{\gamma}}{2},
\end{equation}
where $\theta$ is the angle between the directions of the initial
neutrino propagation and that of the radiated photon. Note that the
structure of eq.(\ref{kinematics}) is very much similar to the
corresponding equation for the case of the $SL\nu$ in matter with
different masses of neutrinos in the initial and final state
\cite{GrigLokhStudTern12}.

To obtain the threshold condition for the $SL\nu$ process one should
consider the system (\ref{eq:conservation}) relative to the vector
$p^{\prime}$. The corresponding exact equation for $p^{\prime}$ is
quadratic and thus can be resolved exactly \cite{LokhovNC2012}. The
resulting threshold condition can be written in the form
\begin{equation}\label{threshold}
   \frac{m^2_{\gamma}+ 2 m_{\gamma}m_{\nu}}{4\widetilde{n}p}<1.
\end{equation}
Note that the condition nontrivially depends on matter density
through $\widetilde{n}$ and $m_\gamma$ given by (\ref{eq:dispersion})
and (\ref{photon dispersion}) respectively.

To meet the requirements on the parameters imposed by the threshold
condition (\ref{threshold}) it is worth to begin with considering the
neutrino mass as its value is very restricted experimentally. To
date, the most conservative upper limit for the neutrino mass which
agrees with existing experiments is set at the order of 1 eV (for the
most recent result from the direct search for neutrino mass refer to
\cite{AsseevPRD11}, for a major review of the field see for instance
\cite{OttWeiRepProgPhys08}). On the other hand, as it was shown in
previous investigations \cite{StuTerPLB05,StuJPA_06_08} that
$SL_{\nu}$ is more efficient for high the matter density. Therefore
we consider the highest densities of matter discussed in the
literature, that are of the order $10^{38-41}~\text{cm}^{-3}$ for
nuclear matter \cite{BelvedNPA12,BowCamZimPRD73}. Indicated values
correspond to the range of the ``density parameter"
$\widetilde{n}\simeq 1-10^{3}~\text{eV}$. Taking also into account
the condition (\ref{n_e&n_n}), from (\ref{photon dispersion}) we
estimate the value of plasmon mass as $m_{\gamma}\simeq
10^{7-8}~\text{eV}$.

With the above scales of the parameters the term that contains
neutrino mass in the numerator in Eq.(\ref{threshold}) is not
important and the threshold condition (7) is reduced to
\begin{equation}\label{threshold_reduced}
    m^2_{\gamma}/4\widetilde{n}p<1.
\end{equation}
That leads, together with the scales of the parameters mentioned
above, to the allowable range for the neutrino momentum $p\gtrsim
10^{12}~\text{eV}$. Thus, even in the case the threshold condition is
accounted for, the room for $SL\nu$ is opened for high energy
neutrinos (see also \cite{KuznMikh11}). The lower the density, the
higher values for the neutrino momentum are needed for the $SL\nu$
process to be opened. The last relation corresponds to the case of
highest matter density among considered, $n\simeq
10^{41}~\text{cm}^{-3}$ (the dependance of the plasmon mass on the
matter density is rather weak).

Note that in spite of the considered ultra-high neutrino energies the
need for accounting for non-local contributions in the
neutrino-matter interactions (emphasized earlier, for instance, in
\cite{KuznMikh07}) is not justified (see also \cite{KuznMikh11}).

\section{The rate and power of the $SL\nu$ process}
Performing calculations similar to those described in \cite{StuTerPLB05} with including additionally the effects of plasma discussed above we arrive at the following expression
for the $SL\nu$ transition rate
\begin{equation}
    \Gamma =\frac{\mu^2}{2\pi}\int\frac{k^2}{\omega}S(k) \, \delta(E-E^{\prime}-\omega) \, dkd\Omega,
\label{eq:Gamma}
\end{equation}
\begin{equation}
    S(k)=(k^2+\omega^2)\left(1-\frac{k-p\cos{\theta}}{p^{\prime}}\cos{\theta}\right)-2\omega{k}
    \left(\cos{\theta}-\frac{k-p\cos{\theta}}{p^{\prime}}\right).
\label{eq:S}
\end{equation}
Here $\mu$ is the neutrino magnetic moment and $\theta$ is the angle
between the direction of the initial neutrino and the plasmon
momentum, $\bf{p}$ and $\bf{k}$ respectively.

The exact evaluation of the integral in (\ref{eq:Gamma}) accounting
for the composite delta function in the presented form is a rather
involved procedure (also has been performed by the authors) and
resulting expression is cumbersome to be presented here. Therefore it
is reasonable to consider several ranges of parameters and obtain
much simpler though less general expressions, that cover the whole
parameters range of interest for astrophysical applications.
Accounting for the threshold condition (\ref{threshold_reduced}) it
is convenient to consider the following three ranges of parameters
and the corresponding three cases.

First we single out the ``near-threshold'' case,
\begin{equation}\label{near threshold}
  m^2_{\gamma}/4\widetilde{n}p \lesssim 1.
\end{equation}
Introducing notation $a=m_{\gamma}^2/4\widetilde{n}p$, the expression
for the process transition rate, that follows from (\ref{near
threshold}), can be written as
\begin{equation}\label{Gamma near threshold}
    \Gamma = 4\mu^2 \, \widetilde{n}^2 p \big((1-a)(1+7a)+4a(1+a)\ln a \big),
\end{equation}
in which one can recognize the estimation from \cite{KuznMikh07}. For
the corresponding expression for the total radiation power
($I=\int{\omega d\Gamma}$) our calculations yield
\begin{equation}\label{I near threshold}
    I =\frac{4}{3} \mu^2 \widetilde{n}^2p^2\big((1-a)(1-5a-8a^2)-12 a^2 \ln a\big).
\end{equation}

The next case we consider is the ``far-from-threshold'' regime,
realized when
\begin{equation}\label{far from threshold}
  m^2_{\gamma}/4\widetilde{n}p \ll 1, \quad \text{or} \quad a \rightarrow 0.
\end{equation}
Since the value for $\widetilde{n}$ is restricted by
$\widetilde{n}\simeq 1-10^{3}~\text{eV}$, the validity of condition
(\ref{far from threshold}) can be provided by the increase of the
neutrino momentum alone, i.e. $p > 10^{12}~\text{eV}$. Thus this
particular case, settled by the condition (\ref{far from threshold})
can be referred to as the case of ``ultra-relativistic'' neutrinos.
In this case the threshold effect is negligible so that the
corresponding result of calculations without account for plasmon mass
should be relevant. Indeed the expressions (\ref{Gamma near
threshold}) and (\ref{I near threshold}) under the condition
(\ref{far from threshold}) are transformed respectively to
\begin{equation}\label{Gamma relativistic}
   \Gamma = 4 \mu^2 \widetilde{n}^2 p,
\end{equation}
\begin{equation}\label{I relativistic}
    I = \frac{4}{3} \mu^2 \widetilde{n}^2p^2,
\end{equation}
which are just results for the $SL\nu$ rate and power at high
neutrino momentum $p$ without inclusion of plasmon mass $m_{\gamma}$
(see  \cite{StuTerPLB05,StuJPA_06_08}).

Now let us consider the case of the $SL\nu$ at the threshold, i.e.
very close to it. In the case the following condition is realized
\begin{equation}\label{close to threshold}
   1-m_{\gamma}^2/4\widetilde{n}p=1-a \ll 1, \quad \text{or} \quad a \rightarrow
   1,
\end{equation}
and the transition rate and radiation power are respectively given by
\begin{equation}\label{Gamma threshold}
   \Gamma = 4\mu^2 \widetilde{n}^2(1-a)\left( (1-a)p +2\widetilde{n} \right),
\end{equation}
\begin{equation}\label{I Threshold}
    I = 4\mu^2 \widetilde{n}^2 p(1-a)\left( (1-a)p +2\widetilde{n} \right).
\end{equation}
The obtained expressions enable us to follow these two
characteristics dependance on the neutrino energy as the parameter
$a$ reaches the threshold, $a \to 1$.  As is expected the rate and
power given by (\ref{Gamma threshold}), (\ref{I Threshold}) go to
zero at the threshold.

\begin{figure}
\begin{center}
  \begin{minipage}{5.5cm}
     \centering
     \includegraphics[scale=0.27]{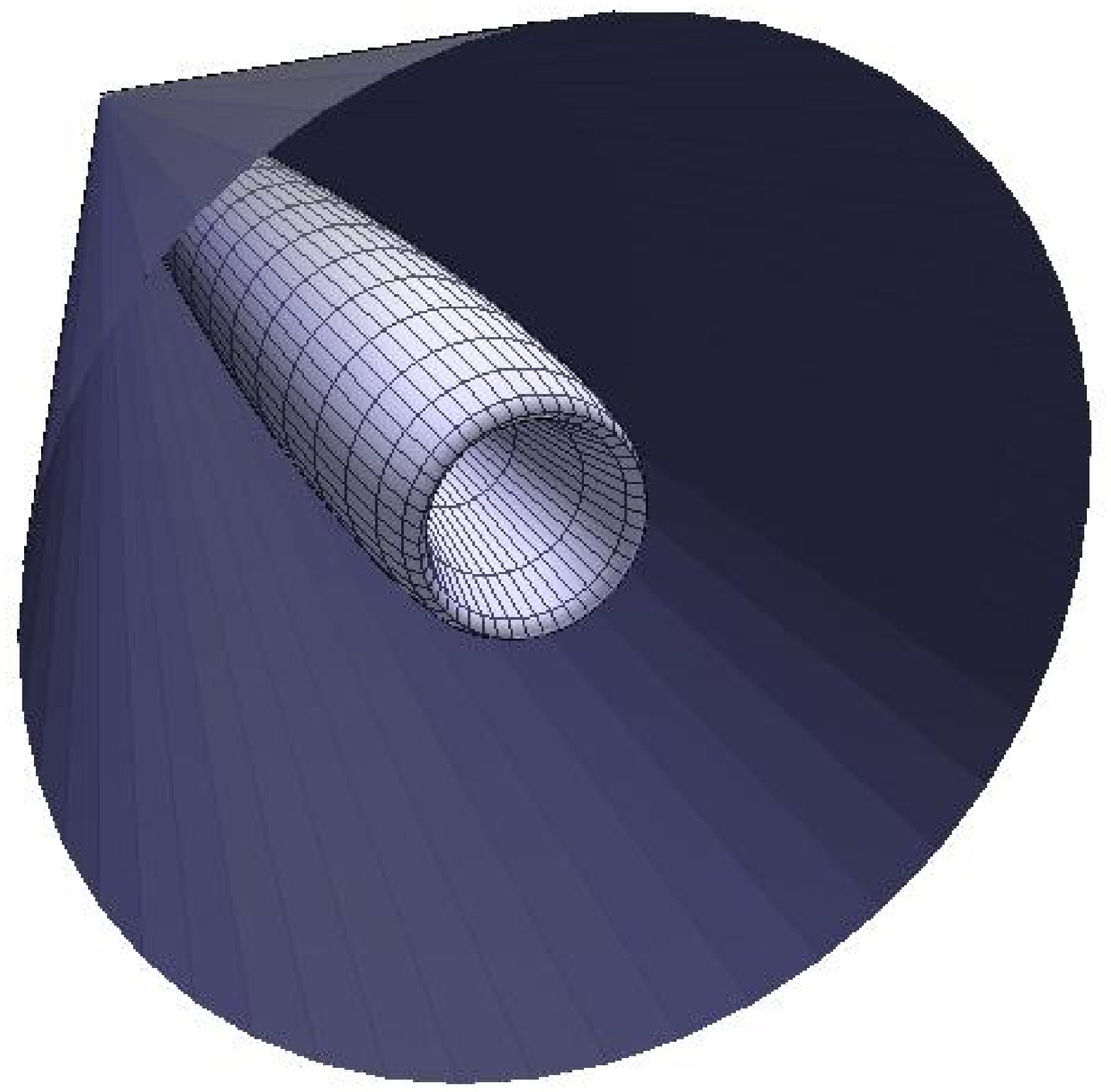}
     \caption{3D representation of the radiation power distribution.}
     \label{dGamma_3D}
  \end{minipage}
  \hspace{2 cm}
  \begin{minipage}{5.5cm}
     \centering
     \includegraphics[scale=0.3]{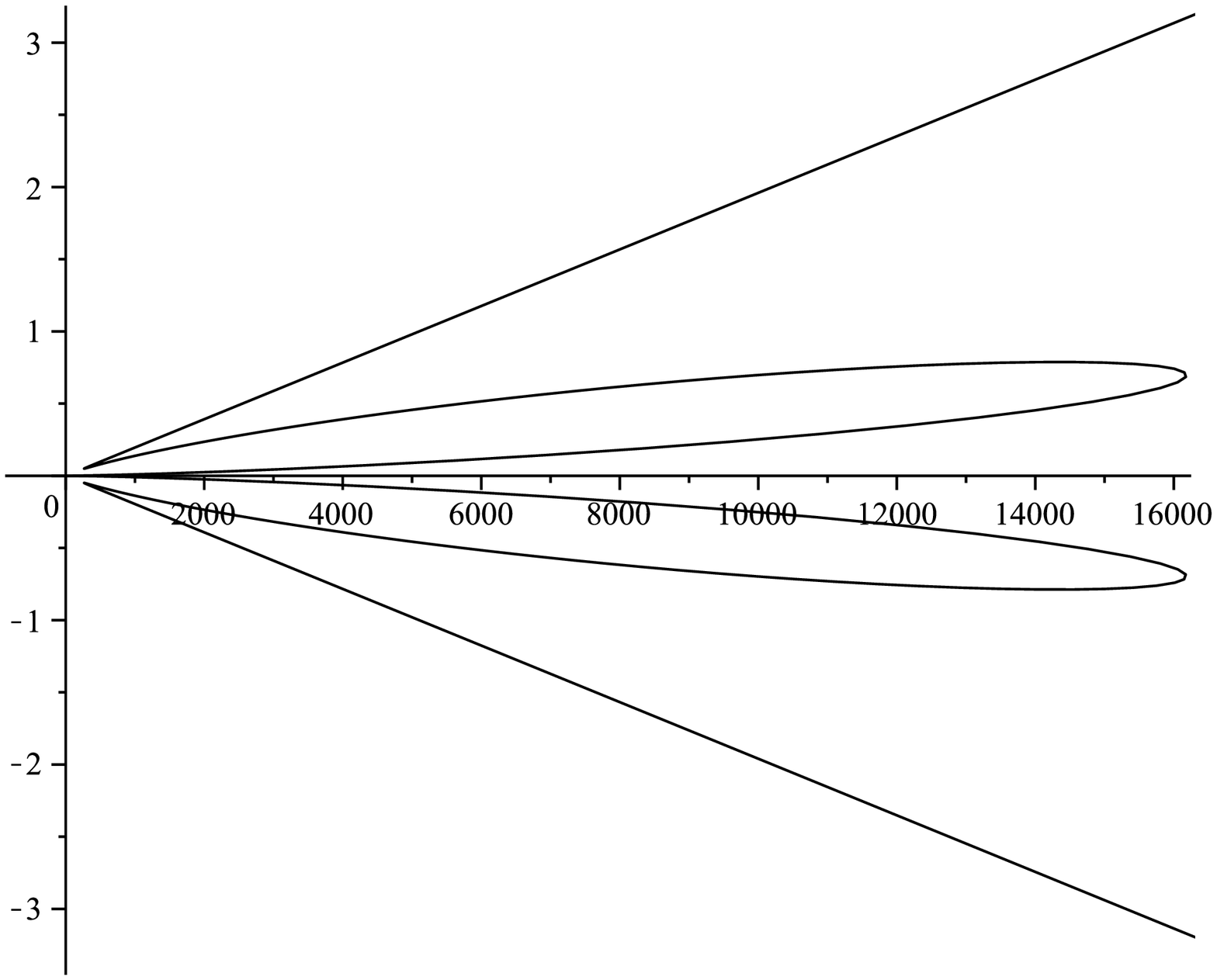}
     \caption{The two-dimensional cut along the symmetry axis. Relative units are used.}
     \label{dGamma_2D}
  \end{minipage}
\end{center}
\end{figure}

The performed above analysis of the $SL\nu$ rate and power enable us
to obtain a fine description for the angular distribution of the
radiation. As follows from (9), (10) the radiation is confined within
the narrow solid angle defined by the relation
\begin{equation}\label{Lim angle}
    \sin \theta \approx \theta \leq \frac{4\widetilde{n}(p + \widetilde{n})  -m^2_{\gamma}}{2p m_{\gamma}}.
\end{equation}
The angular distribution of the $SL\nu$ radiation is presented on
Fig.1 and Fig.2.

Far from the threshold (at high values of the neutrino momentum $p$)
the angular distribution of the $SL\nu$ exhibit the characteristic
that are typical for the radiation produced by a relativistic
particle. As is clearly seen on Fig.1 and Fig.2, the $SL\nu$
radiation is focused in the direction of the initial neutrino
propagation in a very narrow solid angle $\theta \lesssim
2\widetilde{n}/m_{\gamma}$. The radiation distribution has a cup-like
shape at the directions closest to the initial neutrino momentum. A
remarkable feature that one can observe on the base of the $SL\nu$
consideration accounting for the plasma effects is the presence of
the plasmon mass in the angular distribution. Note that the presence
of plasmon mass makes the $SL\nu$ angular distribution analogous to
the angular distribution of on-flight decay of a massive particle. As
it follows from the exact expressions that govern the angular
distribution on Fig.1 and Fig.2, the $SL\nu$ angular distribution
goes to infinity as the angle reaches its boundary value (\ref{Lim
angle}) that corresponds to the outer cone on Fig.1. However the
distribution is integrable and the contribution of the divergent
areas into the total radiation power is negligible as it is common
for kinematics of nuclear reactions (see \cite{Goldansky}).

\section{Conclusions}

We developed a detailed evaluation of the spin light of neutrino in
matter accounting for effects of the emitted plasmon mass. Based on
the exact solution of the modified Dirac equation for the neutrino
wave function in the presence of the background matter the appearance
of the threshold for the considered process is confirmed. The
obtained exact and explicit threshold condition relation exhibit a
rather complicated dependance on the matter density and neutrino
mass. The dependance of the rate and power on the neutrino energy,
matter density and the angular distribution of the $SL\nu$ is
investigated in details. It is shown how the rate and power wash out
when the threshold parameter $a=m_{\gamma}^2/4\widetilde{n}p$
approaching unity (\ref{threshold_reduced}). Within the performed
detailed analysis it is shown that the $SL\nu$ mechanism is
practically insensitive to the emitted plasmon mass for very high
densities of matter (even up to $n=10^{41} cm^{-3}$) in the case of
ultra-high energy neutrinos for a wide range of energies starting
from $E = 1$ TeV. One may expect that the conclusion is of interest
for astrophysical applications of $SL\nu$ radiation mechanism in
light of the recently reported hints of $1\div10$ PeV neutrinos
observed by IceCube \cite{IceCUBE12}.

Although the neutrino fluxes in different astrophysical environments
can be huge, considering possible applications of the $SL\nu$ one
should account for the fact that the mean free path of a neutrino in
neutron matter with densities above $10^{14} \frac{g}{cm^{3}}$
becomes rather short ($\lambda_\nu \sim 10^{-5} cm$) and matter
becomes opaque for neutrinos. Since the neutrino-nucleons cross
section increases with the increasing neutrino energy, the higher the
energy of neutrino the lower should be the density of matter for the
mean free path of neutrino to exceed some characteristic length (for
example, the radius of a neutron star). This is necessary for the
process of the spin light of neutrino to be open.

In the considered case the high density of matter ($n\sim 10^{41}
cm^{-3}$) forbids the propagation of the most energetic neutrinos
with energies above 1 TeV. On the other hand the mass of plasmon in
the matter is high enough to require the neutrino energies within the
TeV scale to get over the threshold. Therefore, these two conditions
(the sufficiently long free path of neutrino and the threshold)
strictly limit the possible parameter space where the $SL\nu$ process
can be effective.

It is worth noting that one can consider the case of a neutrino
moving through dense medium consisting of neutrinos. The studies of
neutrino propagation and oscillation in dense neutrino gases has
started in \cite{SamPRD93} and up to now is under focus of research
(see, for instance, one of the recent papers
\cite{SarTamRafHudJanPRD12} ) The neutrino-neutrino scattering
cross-section is of the same order of magnitude as the
neutrino-nucleon one considered above. Besides the plasmon mass here
becomes irrelevant since the interaction of the emitted photons with
neutrinos of the medium is governed by the neutrino magnetic moment.
In the discussed case the interaction is suppressed as the ratio
$(\frac{\mu_{\nu}^{2}\omega^2}{\alpha})$, where $\alpha$ is the fine
structure constant. Therefore, the threshold condition
(\ref{threshold}) nearly disappears and even the neutrino with rather
low energy (GeV scale or lower) can effectively emit the $SL\nu$.
Such situation could be found during the supernova collapse, for
instance on the edge of neutrinosphere.

In addition there could be other effects that should be considered
within the theory of $SL\nu$. For instance, under certain conditions
the neutrino mixing and oscillations can be important. This situation
can be treated within the approach that was developed for general
kinetic description of relativistic mixed neutrinos that was first
proposed in the pioneering paper \cite{DolgovYF80} and further
studied in \cite{NotRafNPB88,SigRafNPB93,StoPRD87} and adopted for
the neutrino oscillations in the early Universe (see also
\cite{DolPR02}). For this case the neutrino oscillations should be
treated using the neutrino density matrix in flavour space.

However, the latter mentioned effect as well as the newly proposed
possibility for the $SL\nu$ in dense neutrino gases require further
additional detailed investigations.

\section{Acknowledgments}

This study has been partially supported by the Russian Foundation for
Basic Research (grants N. 11-02-01509 and N. 12-02-06833) and the
Ministry of Education and Science of Russia (state contract N.
12.741.11.0180 and projects N. 2012-1.2.1-12-000-1012-1958 and N.
2012-1.1-12-000-1011-6097).

\end{document}